\documentclass[conference, 10pt]{IEEEtran}
\IEEEoverridecommandlockouts

\usepackage[utf8x]{inputenc}
\usepackage[T1]{fontenc}
\usepackage[hidelinks]{hyperref}
\usepackage{babel}
\usepackage{cite}
\usepackage{amsmath,amssymb,amsfonts,mathtools,gensymb}
\usepackage{algorithmic}
\usepackage{graphicx}
\usepackage{textcomp}
\usepackage{xcolor}
\def\BibTeX{{\rm B\kern-.05em{\sc i\kern-.025em b}\kern-.08em
		T\kern-.1667em\lower.7ex\hbox{E}\kern-.125emX}}
\usepackage[final]{microtype}

\usepackage[toc, acronym]{glossaries}
\glsdisablehyper

\usepackage{siunitx}
\usepackage{makecell}
\usepackage{graphicx}
\graphicspath{{figures/}}
\usepackage{enumitem}
\usepackage{siunitx}

\usepackage{tikz}
\usepackage{circuitikz}
\usepackage{pgfplots}
\pgfplotsset{compat=1.16}
\pgfplotsset{compat=newest}
\usepgfplotslibrary{groupplots}
\usepgfplotslibrary{polar}
\usepgfplotslibrary{smithchart}
\usepgfplotslibrary{statistics}
\usepgfplotslibrary{dateplot}
\usepgfplotslibrary{ternary}
\usetikzlibrary{arrows.meta}
\usetikzlibrary{backgrounds}
\usepgfplotslibrary{patchplots}
\usepgfplotslibrary{fillbetween}
\usetikzlibrary{shapes.misc}

\usepackage[nameinlink,capitalize]{cleveref}
\crefname{figure}{Fig.}{Figs.}
\crefname{equation}{}{}
\crefname{section}{Sec.}{Secs.}

\newcommand{\bsf}[1]{\ensuremath{\boldsymbol{\mathsf{#1}}}}
\newcommand{\cov}[1]{\ensuremath{\mathbf{R}_{#1}}}
\newcommand{\hcov}[1]{\ensuremath{\widehat{\mathbf{R}}_{#1}}}
\newcommand{\herm}[1]{\ensuremath{#1^{\mathrm{H}}}}
\newcommand{\trans}[1]{\ensuremath{#1^{\mathrm{T}}}}
\newcommand{\cnumbers}{\ensuremath{\mathbb{C}}}

\newcommand{\complexnormal}[2]{\ensuremath{\mathcal{CN}(#1, #2)}}
\newcommand{\id}[1]{\ensuremath{\mathbf{I}_{#1}}}
\newcommand{\zeros}[1]{\ensuremath{\mathbf{0}_{#1}}}
\newcommand{\expec}[1]{\ensuremath{\mathrm{E}_{#1}}}
\newcommand{\euler}{\ensuremath{\mathrm{e}}}
\newcommand{\im}{\ensuremath{\mathrm{j}}}
\newcommand{\norm}[2]{\ensuremath{\|#1\|_{#2}}}
\newcommand{\Ci}{\ensuremath{\mathrm{C_i}}}
\newcommand{\Si}{\ensuremath{\mathrm{S_i}}}

\newcommand{\tr}{\ensuremath{\mathrm{tr}}}
\newcommand{\snr}{\ensuremath{\mathrm{SNR}}}
\DeclareMathOperator*{\argmin}{arg\,min}
\newcommand{\vs}[1]{\ensuremath{\mathsf{v}_{\mathrm{#1}}}}
\newcommand{\bvs}[1]{\ensuremath{\bsf{v}_{\mathrm{#1}}}}
\newcommand{\is}[1]{\ensuremath{\mathsf{i}_{\mathrm{#1}}}}
\newcommand{\bis}[1]{\ensuremath{\bsf{i}_{\mathrm{#1}}}}
\newcommand{\Z}[1]{\ensuremath{Z_{\mathrm{#1}}}}
\newcommand{\bZ}[1]{\ensuremath{\mathbf{Z}_{\mathrm{#1}}}}
\newcommand{\bsZ}[1]{\ensuremath{\bsf{Z}_{\mathrm{#1}}}}
\newcommand{\bsz}[1]{\ensuremath{\bsf{z}_{\mathrm{#1}}}}
\newcommand{\R}[1]{\ensuremath{R_{\mathrm{#1}}}}
\newcommand{\X}[1]{\ensuremath{X_{\mathrm{#1}}}}
\newcommand{\Pw}[1]{\ensuremath{P_{\mathrm{#1}}}}

\newcommand{\ie}{\textit{i.e. }}

\let\originalleft\left
\let\originalright\right
\renewcommand{\left}{\mathopen{}\mathclose\bgroup\originalleft}
\renewcommand{\right}{\aftergroup\egroup\originalright}

\begin{document}
	\newacronym{simo}{SIMO}{single-input multiple-output}
\newacronym{mimo}{MIMO}{multiple-input multiple-output}
\newacronym{nlos}{NLoS}{non-line-of-sight}
\newacronym{ula}{ULA}{uniform linear array}
\newacronym{bs}{BS}{base station}
\newacronym{snr}{SNR}{signal-to-noise ratio}
\newacronym{lna}{LNA}{low-noise amplifier}
\newacronym{pam}{PAM}{pulse-amplitude modulation}
\newacronym{c}{C}{coherent}
\newacronym{nc}{NC}{noncoherent}
\newacronym{m}{M}{matched}
\newacronym{mm}{MM}{mismatched}
\newacronym{u}{U}{uncoupled}
\newacronym{ser}{SER}{symbol error probability}
\newacronym{ml}{ML}{maximum likelihood}
\newacronym{mrc}{MRC}{maximal ratio combiner}

	\bibliographystyle{IEEEtran-normspace}
	\bstctlcite{IEEEexample:BSTcontrol}

	\title{Coherent and Noncoherent Detection in Dense Arrays: Can We Ignore Mutual Coupling?\\
		\thanks{This work was supported by project MAYTE (PID2022-136512OB-C21) by MICIU/AEI/10.13039/501100011033 and ERDF/EU, grant 2021 SGR 01033 by Departament de Recerca i Universitats de la Generalitat de Catalunya and grant 2023 FI ``Joan Oró'' 00050 by Departament de Recerca i Universitats de la Generalitat de Catalunya and the ESF+.
        L. Sanguinetti was supported in part by the Italian Ministry of Education and Research (MUR) in the framework of the FoReLab Project (Department of Excellence) and in part by the European Union under the Italian National Recovery and Resilience Plan (NRRP) of NextGenerationEU, partnership on ``Telecommunications of the Future'' (PE00000001 -- Program ``RESTART'', Structural Project 6GWINET, Cascade Call SPARKS).}
	}
	\author{
		\IEEEauthorblockN{
			Aniol Martí\IEEEauthorrefmark{1}, Luca Sanguinetti\IEEEauthorrefmark{2}, Jaume Riba\IEEEauthorrefmark{1}, and Meritxell~Lamarca\IEEEauthorrefmark{1}
		}
		\IEEEauthorblockA{
			\textit{\IEEEauthorrefmark{1}Departament de Teoria del Senyal i Comunicacions, Universitat Politècnica de Catalunya (UPC)}\\
		}
		\IEEEauthorblockA{
			\textit{\IEEEauthorrefmark{2}Dipartimento di Ingegneria dell'Informazione, University of Pisa}\\
		}
        \IEEEauthorblockA{
            \{aniol.marti, jaume.riba, meritxell.lamarca\}@upc.edu, luca.sanguinetti@unipi.it
        }
	}
	
	\maketitle
	
	\begin{abstract}
        This paper investigates the impact of mutual coupling on \acrshort{mimo} systems with densely deployed antennas.
        Leveraging multiport communication theory, we analyze both coherent and noncoherent detection approaches in a single-user uplink scenario where the receiver ignores mutual coupling effects.
        Simulation results indicate that while coherent detection is generally more accurate, it is highly sensitive to mismatches in the coupling model, leading to severe performance degradation when antennas are closely spaced, to the point of becoming unusable.
        Noncoherent detection, on the other hand, exhibits a higher error probability but is more robust to coupling model mismatches.
	\end{abstract}
	
	\begin{IEEEkeywords}
        Mutual coupling, holographic MIMO, circuit theory, robust detection, multiport communication theory.
	\end{IEEEkeywords}
	
	\section{Introduction}
        \IEEEPARstart{T}{he increasing} demand for high data rates in modern wireless communication systems has driven significant research into multi-antenna techniques.
        Multiple-input multiple-output (\acrshort{mimo}) systems, and massive \acrshort{mimo} in particular, offer the potential for substantial gains in spectral efficiency and link reliability~\cite[Sec.~4.4]{bjornson_massive_2017}.
        Nevertheless, massive \acrshort{mimo} is not able to cover for all communication requirements envisioned for next generation wireless systems, such as data rates in the order of Tbps or end-to-end latencies below 0.5\,ms~\cite{wang_tutorial_2024}.
        For this reason, new research directions are appearing, with the most prominent ones being large intelligent surfaces~\cite{hu_beyond_2018} and holographic \acrshort{mimo}~\cite{huang_holographic_2020}.
        The latter technology refers to an array with a massive number of densely deployed antennas, which unavoidably results into mutual coupling~\cite[Sec.~8.7]{balanis_antenna_2016}.
        However, most literature on classical and massive \acrshort{mimo} focuses on half-wavelength spaced arrays, where coupling may be neglected~\cite{bjornson_massive_2017}.
        On the other hand, performance analysis of communication systems has mostly relied on abstractions from signal processing and information theory, with the inconvenient of not being consistent with the physical phenomena of the underlying system.
        Fortunately, several approaches to tackle these challenges have been proposed~\cite{migliore_electromagnetics_2008,janaswamy_effect_2002,wallace_mutual_2004,ivrlac_toward_2010,ivrlac_multiport_2014}.
        Among them, a promising one is \textit{multiport communication theory}~\cite{ivrlac_toward_2010,ivrlac_multiport_2014}.
        This method adopts a circuit-theoretic perspective where the multiple-antenna communication system is modeled as a multi-port black box characterized by its impedance matrix.

        In this paper, we leverage multiport communication theory to explore the implications of mutual coupling in \acrshort{mimo} systems with closely spaced antennas.
        Unlike previous works~\cite{janaswamy_effect_2002,wallace_mutual_2004,damico_holographic_2024}, we do not focus on the impact of coupling on \acrshort{snr} or spectral efficiency but rather on symbol detection performance.
        Furthermore, we consider not only coherent detection but also a noncoherent approach, as, to the best of the authors' knowledge, the impact of coupling on the latter has not been investigated.
        In both cases, we assess the robustness of detection against mismatched channel models.
        Specifically, we focus on a single-user uplink scenario where the receiver is unaware of mutual coupling.
        Through numerical simulations, we evidence that while the noncoherent detector generally exhibits a higher error probability, it remains robust to coupling model mismatches due to its inability to exploit phase information.
        In contrast, the coherent detector, though more accurate under ideal conditions, becomes unreliable when antenna spacing is too small, highlighting a fundamental trade-off in system design~\cite[Sec.~5.2.10]{van_trees_detection_2013}.
    
    \section{System Model}
        \label{sec:system_model}
        We consider the uplink of a communication system in \acrfull{nlos} propagation conditions, with a single-antenna transmitter and a receiver with $N$ half-wavelength dipole antennas\footnote{Most communication theory works on mutual coupling consider isotropic radiators~\cite{ivrlac_toward_2010,ivrlac_multiport_2014,yordanov_arrays_2009} or Hertzian dipoles~\cite{marti_asymptotic_2024}, but a more accurate model can be obtained with half-wavelength dipoles~\cite{damico_holographic_2024}.} arranged in a side-by-side configuration as a \acrfull{ula} deployed in the $x$ axis.
        The size of the array is $D$, whereas the separation between antennas is $d=D/(N-1)$ and the wavelength is $\lambda$.
        Hence, the position in polar coordinates of the $n$-th antenna with respect to the origin is $\mathbf{u}_n = \trans{(nd, 0)}$, with $0 \leq n \leq N-1$.
        The discrete baseband representation of the received signal is:
        \begin{equation}
            \label{eq:signal_model}
            \bsf{y} = \bsf{h}\mathsf{x} + \bsf{z},\quad \bsf{y},\bsf{h},\bsf{z}\in\cnumbers^N,
        \end{equation}
        where $\bsf{z}\sim\complexnormal{\zeros{N}}{\cov{\bsf{z}}}$ is additive Gaussian noise and $\mathsf{x}$ is the transmitted symbol.
        Since the noncoherent detector cannot decode phase information, symbols belong to a unipolar \acrfull{pam} constellation $\mathcal{X}=\{0, x_2, \dots, x_M\}$.
        Next, we examine the mutual coupling and spatial correlation models utilized for the characterization of the channel $\bsf{h}$.
        
        \subsection{Multiport Communication}
            \label{sec:multiport}
            \begin{figure}[t]
                \vspace*{-0.7em}
                \centering
                \resizebox{\columnwidth}{!}{%
                    \begin{tikzpicture}[scale=2.54]%
\ifx\dpiclw\undefined\newdimen\dpiclw\fi
\global\def\dpicdraw{\draw[line width=\dpiclw]}
\global\def\dpicstop{;}
\dpiclw=0.8bp
\dpiclw=0.8bp
\dpicdraw (0,-0.393701) rectangle (0.688976,0.393701)\dpicstop
\draw (0.344488,0) node{$\mathbf{Z}_{\mathrm{MT}}$};
\dpicdraw (0,-0.210584)
 --(-0.3,-0.210584)\dpicstop
\dpicdraw (-0.3,-0.210584)
 --(-0.81,-0.210584)\dpicstop
\dpicdraw (-0.804444,-0.210584)
 --(-0.815556,-0.210584)\dpicstop
\dpicdraw (-0.81,-0.210584)
 --(-0.81,-0.11811)\dpicstop
\dpicdraw (-0.81,0) circle (0.0465in)\dpicstop
\draw (-0.81,-0.059055) node{$_-$};
\draw (-0.81,0.059055) node{$_+$};
\dpicdraw (-0.81,0.11811)
 --(-0.81,0.210584)\dpicstop
\draw (-0.92811,0) node[left=-2bp]{$ \mathsf{v}_{\mathrm{G}}$};
\dpicdraw (-0.81,0.205029)
 --(-0.81,0.21614)\dpicstop
\dpicdraw (-0.81,0.210584)
 --(-0.68,0.210584)\dpicstop
\dpicdraw (-0.43,0.210584)
 --(-0.43,0.260584)
 --(-0.68,0.260584)
 --(-0.68,0.160584)
 --(-0.43,0.160584)
 --(-0.43,0.210584)\dpicstop
\dpicdraw (-0.43,0.210584)
 --(-0.3,0.210584)\dpicstop
\draw (-0.555,0.260584) node[above=-2bp]{$ Z_{\mathrm{G}}$};
\filldraw[line width=0bp](-0.205,0.185584)
 --(-0.105,0.210584)
 --(-0.205,0.235584) --cycle\dpicstop
\dpicdraw (-0.127906,0.210584)
 --(-0.205,0.210584)\dpicstop
\draw (-0.166453,0.210584) node[above=-2bp]{$ \mathsf{i}_{\mathrm{T}}$};
\dpicdraw (-0.3,0.210584)
 --(0,0.210584)\dpicstop
\dpicdraw[fill=white](-0.3,-0.210584) circle (0.007874in)\dpicstop
\dpicdraw[fill=white](-0.3,0.210584) circle (0.007874in)\dpicstop
\draw (-0.3,-0.11811) node{$ -$};
\draw (-0.3,0) node{$ \mathsf{v}_{\mathrm{T}}$};
\draw (-0.3,0.11811) node{$ +$};
\dpicdraw (1.240157,-1.043307) rectangle (1.929134,1.043307)\dpicstop
\draw (1.584646,0) node{$\boldsymbol{\mathsf{Z}}_{\mathrm{A}}$};
\dpicdraw (0.688976,0.210584)
 --(0.964567,0.210584)\dpicstop
\filldraw[line width=0bp](1.080316,0.185584)
 --(1.180316,0.210584)
 --(1.080316,0.235584) --cycle\dpicstop
\dpicdraw (1.15741,0.210584)
 --(1.080316,0.210584)\dpicstop
\draw (1.118863,0.210584) node[above=-2bp]{$ \mathsf{i}_{\mathrm{AT}}$};
\dpicdraw (0.964567,0.210584)
 --(1.240157,0.210584)\dpicstop
\dpicdraw (0.688976,-0.210584)
 --(0.964567,-0.210584)\dpicstop
\dpicdraw (0.964567,-0.210584)
 --(1.240157,-0.210584)\dpicstop
\dpicdraw[fill=white](0.964567,0.210584) circle (0.007874in)\dpicstop
\dpicdraw[fill=white](0.964567,-0.210584) circle (0.007874in)\dpicstop
\draw (0.964567,0.11811) node{$ +$};
\draw (0.964567,0) node{$ \mathsf{v}_{\mathrm{AT}}$};
\draw (0.964567,-0.11811) node{$ -$};
\dpicdraw (1.929134,0.797823)
 --(2.154134,0.797823)\dpicstop
\filldraw[line width=0bp](2.112072,0.822823)
 --(2.012072,0.797823)
 --(2.112072,0.772823) --cycle\dpicstop
\dpicdraw (2.034978,0.797823)
 --(2.112072,0.797823)\dpicstop
\draw (2.073525,0.797823) node[above=-2bp]{$ \mathsf{i}_{\mathrm{R},1}$};
\dpicdraw (2.154134,0.797823)
 --(2.298524,0.797823)\dpicstop
\dpicdraw (2.416634,0.797823) circle (0.0465in)\dpicstop
\draw (2.357579,0.797823) node{$_-$};
\draw (2.475689,0.797823) node{$_+$};
\dpicdraw (2.534744,0.797823)
 --(2.679134,0.797823)\dpicstop
\draw (2.416634,0.915933) node[above=-2bp]{$ \mathsf{v}_{\mathrm{EN},1}$};
\dpicdraw (3.204134,0.797823)
 --(3.059744,0.797823)\dpicstop
\dpicdraw (2.941634,0.797823) circle (0.0465in)\dpicstop
\draw (3.000689,0.797823) node{$_-$};
\draw (2.882579,0.797823) node{$_+$};
\dpicdraw (2.823524,0.797823)
 --(2.679134,0.797823)\dpicstop
\draw (2.941634,0.915933) node[above=-2bp]{$ \mathsf{v}_{\mathrm{LNA},1}$};
\dpicdraw (3.198578,0.797823)
 --(3.209689,0.797823)\dpicstop
\dpicdraw (3.204134,0.429597)
 --(3.204134,0.4956)\dpicstop
\dpicdraw (3.204134,0.61371) circle (0.0465in)\dpicstop
\filldraw[line width=0bp](3.229134,0.602293)
 --(3.204134,0.702293)
 --(3.179134,0.602293) --cycle\dpicstop
\dpicdraw (3.204134,0.525127)
 --(3.204134,0.679387)\dpicstop
\dpicdraw (3.204134,0.73182)
 --(3.204134,0.797823)\dpicstop
\draw (3.322244,0.61371) node[right=-2bp]{$ \mathsf{i}_{\mathrm{LNA},1}$};
\dpicdraw (1.929134,0.429597)
 --(2.154134,0.429597)\dpicstop
\dpicdraw (2.154134,0.429597)
 --(2.679134,0.429597)\dpicstop
\dpicdraw (2.679134,0.429597)
 --(3.204134,0.429597)\dpicstop
\dpicdraw (3.198578,0.429597)
 --(3.209689,0.429597)\dpicstop
\dpicdraw[fill=white](2.154134,0.797823) circle (0.007874in)\dpicstop
\dpicdraw[fill=white](2.154134,0.429597) circle (0.007874in)\dpicstop
\draw (2.154134,0.716072) node{$ +$};
\draw (2.154134,0.61371) node{$ \mathsf{v}_{\mathrm{AR},1}$};
\draw (2.154134,0.511348) node{$ -$};
\dpicdraw[fill=white](2.679134,0.797823) circle (0.007874in)\dpicstop
\dpicdraw[fill=white](2.679134,0.429597) circle (0.007874in)\dpicstop
\draw (2.679134,0.716072) node{$ +$};
\draw (2.679134,0.61371) node{$ \mathsf{v}_{\mathrm{R},1}$};
\draw (2.679134,0.511348) node{$ -$};
\dpicdraw (1.929134,-0.429597)
 --(2.154134,-0.429597)\dpicstop
\filldraw[line width=0bp](2.109041,-0.404597)
 --(2.009041,-0.429597)
 --(2.109041,-0.454597) --cycle\dpicstop
\dpicdraw (2.031947,-0.429597)
 --(2.109041,-0.429597)\dpicstop
\draw (2.070494,-0.429597) node[above=-2bp]{$ \mathsf{i}_{\mathrm{R},N}$};
\dpicdraw (2.154134,-0.429597)
 --(2.298524,-0.429597)\dpicstop
\dpicdraw (2.416634,-0.429597) circle (0.0465in)\dpicstop
\draw (2.357579,-0.429597) node{$_-$};
\draw (2.475689,-0.429597) node{$_+$};
\dpicdraw (2.534744,-0.429597)
 --(2.679134,-0.429597)\dpicstop
\draw (2.416634,-0.311487) node[above=-2bp]{$ \mathsf{v}_{\mathrm{EN},N}$};
\dpicdraw (3.204134,-0.429597)
 --(3.059744,-0.429597)\dpicstop
\dpicdraw (2.941634,-0.429597) circle (0.0465in)\dpicstop
\draw (3.000689,-0.429597) node{$_-$};
\draw (2.882579,-0.429597) node{$_+$};
\dpicdraw (2.823524,-0.429597)
 --(2.679134,-0.429597)\dpicstop
\draw (2.941634,-0.311487) node[above=-2bp]{$ \mathsf{v}_{\mathrm{LNA},N}$};
\dpicdraw (3.198578,-0.429597)
 --(3.209689,-0.429597)\dpicstop
\dpicdraw (3.204134,-0.797823)
 --(3.204134,-0.73182)\dpicstop
\dpicdraw (3.204134,-0.61371) circle (0.0465in)\dpicstop
\filldraw[line width=0bp](3.229134,-0.625127)
 --(3.204134,-0.525127)
 --(3.179134,-0.625127) --cycle\dpicstop
\dpicdraw (3.204134,-0.702293)
 --(3.204134,-0.548034)\dpicstop
\dpicdraw (3.204134,-0.4956)
 --(3.204134,-0.429597)\dpicstop
\draw (3.322244,-0.61371) node[right=-2bp]{$ \mathsf{i}_{\mathrm{LNA},N}$};
\dpicdraw (1.929134,-0.797823)
 --(2.154134,-0.797823)\dpicstop
\dpicdraw (2.154134,-0.797823)
 --(2.679134,-0.797823)\dpicstop
\dpicdraw (2.679134,-0.797823)
 --(3.204134,-0.797823)\dpicstop
\dpicdraw (3.198578,-0.797823)
 --(3.209689,-0.797823)\dpicstop
\dpicdraw[fill=white](2.154134,-0.429597) circle (0.007874in)\dpicstop
\dpicdraw[fill=white](2.154134,-0.797823) circle (0.007874in)\dpicstop
\draw (2.154134,-0.511348) node{$ +$};
\draw (2.154134,-0.61371) node{$ \mathsf{v}_{\mathrm{AR},N}$};
\draw (2.154134,-0.716072) node{$ -$};
\dpicdraw[fill=white](2.679134,-0.429597) circle (0.007874in)\dpicstop
\dpicdraw[fill=white](2.679134,-0.797823) circle (0.007874in)\dpicstop
\draw (2.679134,-0.511348) node{$ +$};
\draw (2.679134,-0.61371) node{$ \mathsf{v}_{\mathrm{R},N}$};
\draw (2.679134,-0.716072) node{$ -$};
\dpicdraw[fill=black](2.679134,0.109113) circle (0.007874in)\dpicstop
\dpicdraw[fill=black](2.679134,0) circle (0.007874in)\dpicstop
\dpicdraw[fill=black](2.679134,-0.109113) circle (0.007874in)\dpicstop
\dpicdraw (3.204134,0.797823)
 --(3.879134,0.797823)\dpicstop
\dpicdraw (3.873578,0.797823)
 --(3.884689,0.797823)\dpicstop
\dpicdraw (3.879134,0.797823)
 --(3.879134,0.73871)\dpicstop
\dpicdraw (3.879134,0.48871)
 --(3.929134,0.48871)
 --(3.929134,0.73871)
 --(3.829134,0.73871)
 --(3.829134,0.48871)
 --(3.879134,0.48871)\dpicstop
\dpicdraw (3.879134,0.48871)
 --(3.879134,0.429597)\dpicstop
\draw (3.929134,0.61371) node[right=-2bp]{$ Z_{\mathrm{L}}$};
\draw (4.312205,0.77119) node{$ +$};
\draw (4.312205,0.61371) node{$ \mathsf{v}_{\mathrm{L},1}$};
\draw (4.312205,0.45623) node{$ -$};
\dpicdraw (3.879134,0.435153)
 --(3.879134,0.424041)\dpicstop
\dpicdraw (3.879134,0.429597)
 --(3.204134,0.429597)\dpicstop
\dpicdraw (3.204134,-0.429597)
 --(3.879134,-0.429597)\dpicstop
\dpicdraw (3.873578,-0.429597)
 --(3.884689,-0.429597)\dpicstop
\dpicdraw (3.879134,-0.429597)
 --(3.879134,-0.48871)\dpicstop
\dpicdraw (3.879134,-0.73871)
 --(3.929134,-0.73871)
 --(3.929134,-0.48871)
 --(3.829134,-0.48871)
 --(3.829134,-0.73871)
 --(3.879134,-0.73871)\dpicstop
\dpicdraw (3.879134,-0.73871)
 --(3.879134,-0.797823)\dpicstop
\draw (3.929134,-0.61371) node[right=-2bp]{$ Z_{\mathrm{L}}$};
\draw (4.312205,-0.45623) node{$ +$};
\draw (4.312205,-0.61371) node{$ \mathsf{v}_{\mathrm{L},N}$};
\draw (4.312205,-0.77119) node{$ -$};
\dpicdraw (3.879134,-0.792268)
 --(3.879134,-0.803379)\dpicstop
\dpicdraw (3.879134,-0.797823)
 --(3.204134,-0.797823)\dpicstop
\end{tikzpicture}%
                }
                \caption{Circuit model of a \acrshort{simo} communication system.}
                \label{fig:uplink_multiport}
            \end{figure}
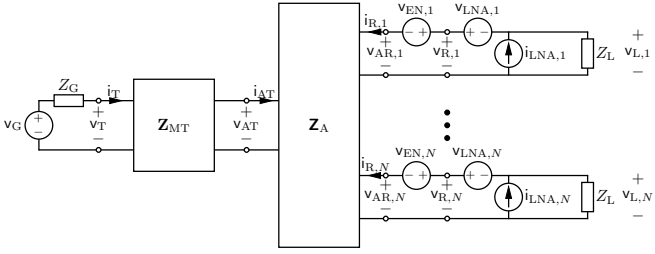

            Mutual coupling is modeled using multiport communication theory, a framework that involves a circuit-theoretic approach of the wireless channel~\cite{ivrlac_toward_2010,ivrlac_multiport_2014}.
            The multiport model for the system under consideration is depicted in~\cref{fig:uplink_multiport}, and it consists of four parts:
            \begin{enumerate}[label=\textit{\arabic*)}]
                \item \textit{Signal generation:} The signal generated by the transmitter is modeled by a voltage source with complex envelope $\vs{G}$, in series with an impedance $\Z{G} = \R{G}+\im\X{G}$.
                The total average available power is:
                \begin{equation}
                    \label{eq:available_power}
                    \Pw{a} = \expec{}\Big[\frac{|\vs{G}|^2}{4\R{G}}\Big].
                \end{equation}
                \item \textit{Impedance matching:} Impedance matching networks are used to optimize power transfer to the transmitter antennas or to enhance the \acrfull{snr} at the receiver.
                However, when the number of antennas is large, designing optimal matching networks is quite complex~\cite{mezghani_reincorporating_2023}, therefore we only consider the transmitter matching network~\cite{damico_holographic_2024}:
                \begin{equation}
                    \bZ{MT} = \begin{pmatrix}
                        -\im\X{G} & -\im\sqrt{\R{G}\Re({\Z{AT}})}\\
                        -\im\sqrt{\R{G}\Re({\Z{AT}})} & -\im\Im(\Z{AT})
                    \end{pmatrix}.
                \end{equation}
                \item \textit{Antenna coupling:} Mutual coupling between antennas is described by the impedance matrix:
                \begin{equation}
                    \bsZ{A} = \begin{pmatrix}
                        \Z{AT} & \zeros{}\\
                        \bsz{{ART}} & \bZ{R}
                    \end{pmatrix} \in \cnumbers^{(N+1)\times(N+1)},
                \end{equation}
                where $\Z{AT}$ is the transmitting antenna impedance, $\bsz{ART}$ models the inter-array coupling from the transmitter to the receiver and $\bZ{R}$ the intra-array coupling at the receiver.
                The classical uncoupled model is obtained by setting $\bZ{R}=(\R{r} + \im\X{A})\id{N}$, with $\R{r}$ the radiation resistance and $\X{A}$ the antenna reactance~\cite[Sec.~2.13]{balanis_antenna_2016}.
                The coupling from the receiver to the transmitter is assumed to be negligible (\ie unilateral approximation)~\cite{marti_asymptotic_2024}.
                \item \textit{Noise:} The voltage sources $\bvs{EN} = \trans{(\vs{EN}{}_{,1},\dots,\vs{EN}{}_{,N})}$ account for the extrinsic noise, which comes from background radiation and is received by the antennas.
                A usual model for extrinsic noise is $\bvs{EN}\sim\complexnormal{\zeros{N}}{\cov{\mathrm{EN}}}$, where the covariance matrix is:
                \begin{equation}
                    \cov{\mathrm{EN}} = 4k_{\mathrm{B}}T_{\mathrm{A}}B_{\mathrm{W}} \Re(\bZ{R}),
                \end{equation}
                where $k_{\mathrm{B}}$ is the Boltzmann constant, $T_{\mathrm{A}}$ the noise temperature of the antennas and $B_{\mathrm{W}}$ the equivalent noise bandwidth~\cite{ivrlac_toward_2010}.
                Intrinsic noise, on the other hand, originates from the components connected after the antennas such as \acrfullpl{lna}.
                Intrinsic noise can be modeled with the current and voltage vectors $\bis{LNA}\sim\complexnormal{\zeros{N}}{\sigma_i^2\id{N}}$ and $\bvs{LNA}\sim\complexnormal{\zeros{N}}{\R{N}^2\sigma_i^2\id{N}}$, where $\R{N}$ is the noise resistance of the \acrshortpl{lna} given by the manufacturer.
                Although extrinsic and intrinsic noise are assumed to be uncorrelated, the voltage and current in the same amplifier are not:
                \begin{equation}
                    \expec{}[\bvs{LNA}\herm{\bis{LNA}}] = \rho\R{N}\sigma_i^2\id{N},
                \end{equation}
                where $\rho$ is the noise correlation coefficient.
            \end{enumerate}

            Letting $\bvs{L} = \trans{(\vs{L}{}_{,1},\dots,\vs{L}{}_{,N})}$ and taking into account that the circuit model is linear, the input-output relation must be of the form:
            \begin{equation}
                \label{eq:input_output}
                \bvs{L} = \bsf{d}\vs{G} + \bsf{n},
            \end{equation}
            where, by superposition, $\bsf{d}\vs{G}$ is the signal component and $\bsf{n}\sim\complexnormal{\zeros{N}}{\cov{\bsf{n}}}$ is the noise component.
            The former can be obtained by setting $\bvs{EN} = \bvs{LNA} = \bis{LNA} = \zeros{}$ and combining $\bZ{MT}$ and $\bZ{A}$ into a single multiport:
            \begin{equation}
                \begin{pmatrix}
                    \vs{T}\\
                    \bvs{L}
                \end{pmatrix} =
                \begin{pmatrix}
                    \Z{T} & 0\\
                    \bsz{RT} & \bZ{R}
                \end{pmatrix}\begin{pmatrix}
                    \is{T}\\
                    \bis{R}
                \end{pmatrix},
            \end{equation}
            where $\Z{T}$ and $\bsz{RT}$ are obtained from circuit analysis as:
            \begin{align}
                \Z{T} &= [\bZ{MT}]_{1,1} - \frac{[\bZ{MT}]_{1,2}^2}{\Z{AT}+[\bZ{MT}]_{2,2}} = \R{G}-\im\X{G},\\
                \bsz{RT} &= \frac{[\bZ{MT}]_{2,1}}{\Z{AT}+[\bZ{MT}]_{2,2}}\bsz{ART} = -\im\Big(\frac{\R{G}}{\Re(\Z{AT})}\Big)^{\frac{1}{2}}\bsz{ART}.
            \end{align}
            From Ohm's law, $\bis{R}=\frac{\vs{L}}{\Z{L}}$ and $\is{T}=\frac{\vs{G}-\vs{T}}{\Z{G}}$, thus:
            \begin{equation}
                \label{eq:channel}
                \bsf{d} = \frac{-\im}{2\sqrt{\R{G}\Re(\Z{AT})}}\,\mathbf{Q}\,\bsz{ART},
            \end{equation}
            where $\mathbf{Q}$ is an auxiliary matrix defined as:
            \begin{equation}
                \mathbf{Q} = \Z{L}(\Z{L}\id{N}+\bZ{R})^{-1}.
            \end{equation}
            The noise term is obtained from circuit analysis after setting the input to zero (\ie $\vs{G}=0$):
            \begin{equation}
                \bsf{n} = \mathbf{Q}(\bvs{EN}-\bvs{LNA}+\bZ{R}\bis{LNA}).
            \end{equation}
            As it is a zero-mean Gaussian random variable, it is fully characterized by its covariance matrix:
            \begin{align}
                \cov{\bsf{n}} &= \mathbf{Q}(\cov{\mathrm{EN}} + \mathbf{U}_{\mathrm{LNA}})\herm{\mathbf{Q}},\label{eq:noise_correlation}\\
                \mathbf{U}_{\mathrm{LNA}} &= \sigma_i^2(\R{N}^2\id{N} + \bZ{R}\herm{\bZ{R}} - 2\R{N}\Re(\rho^*\bZ{R})).
            \end{align}

            \subsection{Inter-Array Coupling}
                Inter-array coupling $\bsz{ART}$ depends on the propagation conditions between transmitter and receiver.
                In a \acrshort{nlos} environment with $L$ scatterers in the radiative near field, inter-array coupling can be modeled as~\cite[Sec.~2.6]{ivrlac_multiport_2014,bjornson_massive_2017}:
                \begin{equation}
                    \bsz{ART} = \im\R{r}\sum_{i=1}^{L}\alpha_i\mathbf{a}_i,
                \end{equation}
                where $\alpha_i \in \cnumbers$ is the complex gain of the $i$-th scatterer and
                \begin{equation}
                    \mathbf{a}_i = \trans{\left(\frac{\euler^{-\im k\norm{\mathbf{s}_{i}-\mathbf{u}_0}{}}}{k\norm{\mathbf{s}_{i}-\mathbf{u}_0}{}}, \dots, \frac{\euler^{-\im k\norm{\mathbf{s}_{i}-\mathbf{u}_{N-1}}{}}}{k\norm{\mathbf{s}_{i}-\mathbf{u}_{N-1}}{}}\right)},
                \end{equation}
                with $k=2\pi/\lambda$, is the array response vector for the wave arriving from the scatterer located at $\mathbf{s}_{i} = \trans{(r_{i}, \theta_{i})}$.
                When $L$ is sufficiently large, $\bsz{ART}$ follows a complex Gaussian distribution $\bsz{ART}\sim\complexnormal{\zeros{N}}{\cov{\mathrm{ART}}}$ with covariance:
                \begin{equation}
                    \cov{\mathrm{ART}} = \expec{}[\bsz{ART}\herm{\bsz{ART}}] = \R{r}^2\sum_{i=1}^{L}\beta_i\mathbf{a}_i\herm{\mathbf{a}_i},
                \end{equation}
                where $\beta_i = \expec{}[|\alpha_i|^2]$.
                This model assumes spherical wavefronts and point antennas, hence it is valid in the radiative near field and the far field~\cite{dong_near-field_2022,bjornson_primer_2021}.
                        
            \subsection{Intra-Array Coupling}
                \label{sec:intra-array_coupling}
                When modeling intra-array coupling, the actual shape and position of the antennas must be taken into account.
                In the case of side-by-side center-fed half-wavelength dipoles, the self-impedance (\ie the diagonal of $\bZ{R}$) is $\R{r} +\im \X{A} = 73 + \im42.5\,\Omega$, whereas the mutual impedances can be computed as~\cite[Eq.~8.68]{balanis_antenna_2016}:
                \begin{align}
                    [\Re(\bZ{R})]_{p,q} = \frac{\eta}{4\pi}[2\Ci(u_{p,q})-\Ci(v_{p,q})-\Ci(w_{p,q})],\label{eq:real_coupling}\\
                    [\Im(\bZ{R})]_{p,q} = -\frac{\eta}{4\pi}[2\Si(u_{p,q})-\Si(v_{p,q})-\Si(w_{p,q})],
                \end{align}
                where $\Si(\cdot)$ and $\Ci(\cdot)$ are the sine and cosine integrals~\cite[Appendix~III]{balanis_antenna_2016}, and
                \begin{align}
                    u_{p,q} &= kd|p-q|,\\
                    v_{p,q} &= k(\sqrt{d^2(p-q)^2+l^2}+l),\\
                    w_{p,q} &= k(\sqrt{d^2(p-q)^2+l^2}-l),
                \end{align}
                with $l=\lambda/2$ the dipole length.

\begin{figure}[t]
                    \vspace*{-0.5em}
    				\centering
    				\resizebox{\columnwidth}{!}{%
    					\input{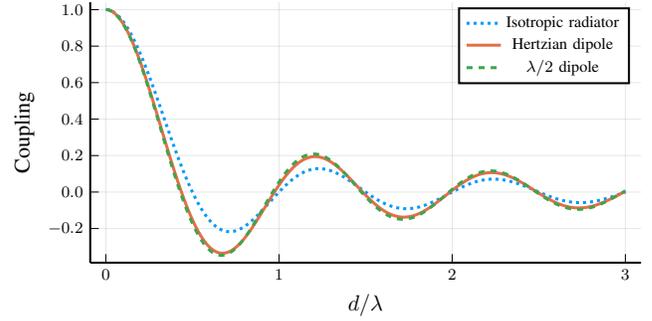}
    				}
    				\caption{Mutual coupling between two isotropic radiators, Hertzian dipoles or half-wavelength dipoles at different distances.}
    				\label{fig:coupling}
                \end{figure}

                The real part of the coupling function between two half-wavelength dipoles (\ie \cref{eq:real_coupling} for an arbitrary separation and normalized to one) is depicted in~\cref{fig:coupling}, together with the coupling functions for two Hertzian dipoles~\cite[Eq.~(14)]{marti_asymptotic_2024} and two isotropic radiators~\cite[Eq.~(7)]{yordanov_arrays_2009}.
                It can be observed that both Hertzian and half-wavelength dipoles exhibit practically the same mutual coupling effect, yet a scaling factor due to their different radiation resistance should be taken into account.
                On its turn, the coupling of two isotropic radiators follows the pattern of a cardinal sine, which is qualitatively very similar to that of the dipoles, as it was previously noted in~\cite{yordanov_arrays_2009}.

                In the sequel, we show that the mutual coupling function is tightly related to the error probability of a coherent communication system.
                
            \subsection{Channel Model}
                In order to obtain the dimensionless expression in~\cref{eq:signal_model} from~\cref{eq:input_output} we consider:
                \begin{equation}
                    \frac{\bvs{L}}{\sqrt{c}} = \bsf{d}\frac{\vs{G}}{\sqrt{c}} + \frac{\bsf{n}}{\sqrt{c}},
                \end{equation}
                where $c$ is a constant measured in $\mathrm{V}^2$.
                Then, letting $\bsf{y} = \bvs{L}/\sqrt{c}$, $\bsf{h} = \bsf{d}$, $\mathsf{x} = \vs{G}/\sqrt{c}$ and $\bsf{z} = \bsf{n}/\sqrt{c}$ we obtain the \acrshort{mimo} system input-output relation, with the following second-order constraints:
                \begin{align}
                    \expec{}[|\mathsf{x}|^2] &= 4\R{G}\Pw{a}/c,\label{eq:const_power}\\
                    \cov{\bsf{z}} &= \cov{\bsf{n}}/c,\label{eq:equivalent_noise_correlation}\\
                    \cov{\bsf{h}} &= \frac{1}{4\R{G}\Re(\Z{AT})}\mathbf{Q}\cov{\mathrm{ART}}\herm{\mathbf{Q}},\label{eq:channel_correlation}
                \end{align}
                where~\cref{eq:const_power} follows from~\cref{eq:available_power} and the power matching network in the transmitter.
                As $c$ is arbitrary, we assume $c=\qty{1}{\V^2}$.

            \section{Receiver Structure}
                The main goal of this paper is to characterize the impact of ignoring antenna coupling in the detection stage.
                Both coherent and noncoherent approaches are considered herein.
                Following state of the art literature on high frequency \acrshort{mimo} communications~\cite{bacci_spherical_2023,marti_harnessing_2025}, the \acrfull{bs} implements a model-based acquisition of the channel realization and correlation.
                This means that, instead of estimating them directly, the receiver estimates their parameters and constructs $\mathbf{h}$ and $\cov{\bsf{h}}$ from them, employing the models~\cref{eq:channel,eq:channel_correlation}.
                While model-agnostic methods rely on unstructured channel estimation typically based on pilot transmissions in the uplink, the use of model-based methods involving the estimation of environment parameters with physical sense usually leads to better estimation quality, specially in the low \acrshort{snr} regime.
                Nevertheless, it is important to verify in practice whether or not the precision in modeling can be relaxed to simplify the receiver designs.

                The \acrfull{ser} is defined as:
                \begin{equation}
                        P_\epsilon = \frac{1}{M}\sum_{x\in\mathcal{X}}\Pr[\widehat{x}(\bsf{y}) \neq x | \mathsf{x} = x],
                \end{equation}
                where $\widehat{x}(\bsf{y})$ is the output of a symbol detector applied to $\bsf{y}$.

                In~\cref{sec:numerical_results}, we evaluate the performance of the optimal \acrfull{c} and optimal one-shot \acrfull{nc} detectors in three different scenarios: a \acrfull{m} detector that employs the correct signal model, a \acrfull{mm} detector that accounts for the scatterers model and location but ignores mutual coupling (\ie $\bZ{R} = (\R{r}+\im \X{A})\id{N}$), and a detector designed for an artificial \acrfull{u} scenario where the signal is generated and detected with $\bZ{R} = (\R{r}+\im \X{A})\id{N}$.
                                
                \subsection{Noncoherent Detector}
                    For a one-shot receiver unknowing of instantaneous channel realizations but only its statistics, the optimal symbol decision is given by the unconditional \acrlong{ml} detector\footnote{Under uncorrelated Rayleigh fading, this receiver is often referred as \textit{average channel energy detector}~\cite{jing_design_2016}.}~\cite{vila-insa_quadratic_2024}:
                    \begin{equation}
                        \widehat{x}_{\mathrm{NC}} = \argmin_{x\in\mathcal{X}} \herm{\mathbf{y}}\cov{\bsf{y}|x}^{-1}\mathbf{y} + \log|\cov{\bsf{y}|x}|,
                    \end{equation}
                    where
                    \begin{equation}
                        \label{eq:ml_correlation}
                        \cov{\bsf{y}|x} = |x|^2\cov{\bsf{h}} + \cov{\bsf{z}},
                    \end{equation}
                    with $\cov{\bsf{h}}$ and $\cov{\bsf{z}}$ given by~\cref{eq:channel_correlation,eq:equivalent_noise_correlation}.
                    The average \acrshort{snr} at the receiver is:
                    \begin{equation}
                        \snr = \frac{\expec{}[|\mathsf{x}|^2\herm{\bsf{h}}\bsf{h}]}{\expec{}[\herm{\bsf{z}}\bsf{z}]} = \frac{4\R{G}\Pw{a}\tr(\cov{\bsf{h}})}{\tr(\cov{\bsf{z}})}.
                    \end{equation}
                    If the detector is designed disregarding mutual coupling, the correlation matrix employed by the detector is not~\cref{eq:ml_correlation} but:
                    \begin{equation}
                        \hcov{\bsf{y}|x} = \gamma_1|x|^2\cov{\mathrm{ART}} + \gamma_2\id{N},
                    \end{equation}
                    with $\gamma_1$ and $\gamma_2$ power scaling factors obtained from the multiport model in~\cref{sec:system_model} with $\bZ{R}=(\R{r}+\im \X{A})\id{N}$.
                    
                \subsection{Coherent Detector}
                    Instead, if full knowledge of every single channel realization is available at the receiver, the optimal decision is obtained with the \acrlong{mrc}~\cite[Ch.~3]{tse_fundamentals_2005}:
                    \begin{equation}
                        \widehat{x}_{\mathrm{C}} = \argmin_{x\in\mathcal{X}} \left\lvert\frac{\Re[\herm{\mathbf{h}}\cov{\bsf{z}}^{-1}\mathbf{y}]}{\herm{\mathbf{h}}\cov{\bsf{z}}^{-1}\mathbf{h}}-x\right\rvert.
                    \end{equation}
                    Like in the noncoherent case, if it is assumed that the \acrshort{bs} ignores mutual coupling, the detector operates in a mismatched mode with the channel and noise correlation given by:
                    \begin{align}
                        &\widehat{\mathbf{h}} = \sqrt{\gamma_1}\mathbf{z}_{\mathrm{ART}},\\
                        &\hcov{\bsf{z}} = \gamma_2\id{N},
                    \end{align}
                    where $\gamma_1$ and $\gamma_2$ are the same scaling factors as before.
                
            \section{Numerical Results}
                \label{sec:numerical_results}
                \begin{table*}[t!]
                    \renewcommand{\arraystretch}{1.3}
                    \centering
                    \caption{Summary of simulation parameters.}
                    \begin{tabular}{c|c|c|c}
                        \textbf{Parameter} & \textbf{Value} & \textbf{Parameter} & \textbf{Value}\\
                        \hline
                        Carrier frequency & $f = \qty{30}{\GHz}$ & Variance of the current noise source & $\sigma^2_i = 2 k_{\mathrm{B}} B_{\mathrm{W}} T_{\mathrm{A}}/\R{N}$\\
                        \hline
                        Bandwidth  & $B_{\mathrm{W}} = \qty{20}{\MHz}$ & Antenna impedance & $\Z{A} = 73+\im42.5\,\Omega$ \\
                        \hline
                        Amplifier and load impedance  & $\Z{G}= \Z{L} = 186-\im31.6\,\Omega$ & Complex correlation coefficient& $\rho = 0.2730+\im0.1793$ \\
                        \hline
                        Noise temperature of antennas  & $T_{\mathrm{A}}= \qty{290}{\K}$ & \acrshort{lna} resistance & $\R{N} = \qty{5}{\ohm}$ \\
                        \hline
                    \end{tabular}
                    \label{tab:simulation_parameters}
                \end{table*}
      
                In order to assess the performance degradation of a receiver unaware of antenna mutual coupling, we consider a scenario with a single user located at $\mathbf{r} = \trans{(r, \theta)} = \trans{(\qty{25}{\m}, \qty{-30}{\degree})}$, surrounded by $L=20$ scatterers uniformly sampled from a circle centered at the user, with radius $r_{\mathrm{c}} = \qty{3}{\m}$, unless stated otherwise.
                Symbols are transmitted from a $4$-\acrshort{pam} unipolar constellation and the average \acrshort{snr} is $\qty{5}{\decibel}$.
                The remaining parameters are summarized in~\cref{tab:simulation_parameters}, with values from~\cite{lehmeyer_lna_2015, damico_holographic_2024} and references therein.

\begin{figure}[t]
                    \vspace*{-0.3em}
    				\centering
    				\resizebox{\columnwidth}{!}{%
    					\input{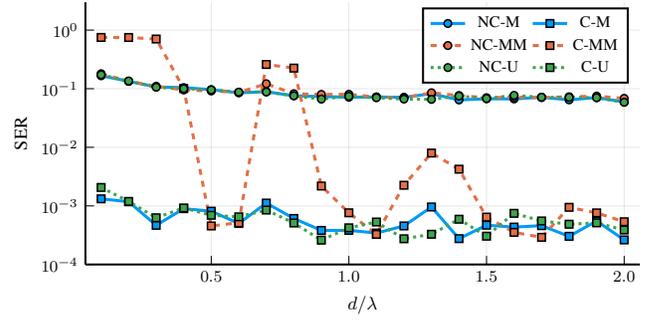}
    				}
    				\caption{Error probability of the matched (M) and mismatched (MM), coherent (C) and noncoherent (NC), and uncoupled (U) detectors as a function of element separation for $N=128$.}
    				\label{fig:ser_coupled_d}
                \end{figure}

                In general, the noncoherent detector exhibits a much higher error probability, but it is more robust to mismatches in coupling.
                This is consistent with the well-known trade-off between robustness and optimality.
                The difference in operation can be observed in~\cref{fig:ser_coupled_d}, where the coherent mismatched detector \acrshort{ser} follows the same pattern as the coupling function in~\cref{sec:intra-array_coupling}.
                On the other hand, the energy detector performance is almost unaffected by the mismatch.
                
\begin{figure}[t]
    				\centering
    				\resizebox{\columnwidth}{!}{%
    					\input{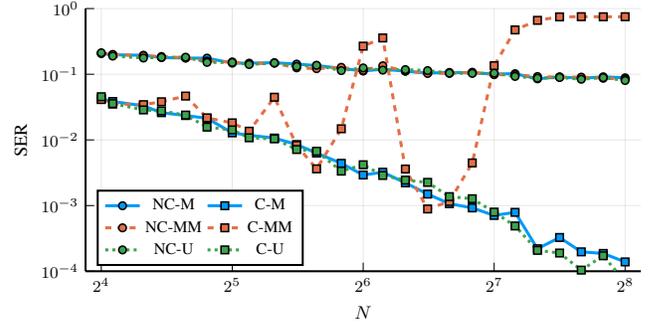}
    				}
    				\caption{Error probability of the matched (M) and mismatched (MM), coherent (C) and noncoherent (NC), and uncoupled (U) detectors in terms of $N$ for a fixed aperture $D=\qty{0.5}{\m}$.}
    				\label{fig:ser_coupled_n}
                    \vspace*{-1.5ex}
                \end{figure}

                A similar behavior is illustrated in~\cref{fig:ser_coupled_n}.
                When the array aperture is fixed but the number of antennas increases, the inter-element distance decreases and the coupling also increases.
                In this situation, the coherent mismatched receiver performance rapidly decays and becomes unusable.
                On the contrary, the noncoherent detector error probability is robust to the mismatch, and it even improves with $N$.
                In the simulated example, the noncoherent detector outperforms the coherent for $N\geq128$.                

                It should also be remarked that the \acrshort{ser} of a receiver, either coherent or noncoherent, that perfectly knows the effect of mutual coupling is practically equal to that of an uncoupled scenario, as can be observed by comparing the solid blue and dotted green lines in~\cref{fig:ser_coupled_d} and~\cref{fig:ser_coupled_n}.

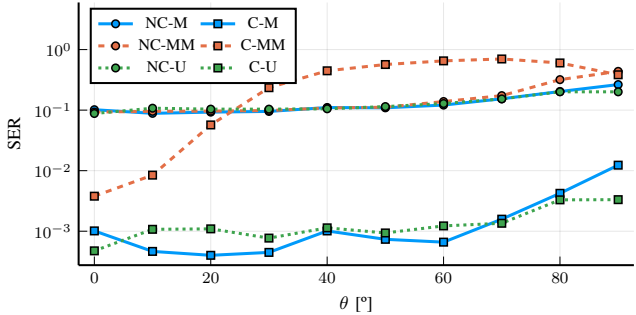
\begin{figure}[t]
    				\centering
    				\resizebox{\columnwidth}{!}{%

\begin{tikzpicture}[/tikz/background rectangle/.style={fill={rgb,1:red,1.0;green,1.0;blue,1.0}, fill opacity={1.0}, draw opacity={1.0}}, show background rectangle]
\begin{axis}[point meta max={nan}, point meta min={nan}, legend cell align={left}, legend columns={2}, title={}, title style={at={{(0.5,1)}}, anchor={south}, font={{\fontsize{14 pt}{18.2 pt}\selectfont}}, color={rgb,1:red,0.0;green,0.0;blue,0.0}, draw opacity={1.0}, rotate={0.0}, align={center}}, legend style={color={rgb,1:red,0.0;green,0.0;blue,0.0}, draw opacity={1.0}, line width={1}, solid, fill={rgb,1:red,1.0;green,1.0;blue,1.0}, fill opacity={1.0}, text opacity={1.0}, font={{\fontsize{8 pt}{10.4 pt}\selectfont}}, text={rgb,1:red,0.0;green,0.0;blue,0.0}, cells={anchor={center}}, at={(0.02, 0.98)}, anchor={north west}}, axis background/.style={fill={rgb,1:red,1.0;green,1.0;blue,1.0}, opacity={1.0}}, anchor={north west}, xshift={1.0mm}, yshift={-1.0mm}, width={112.3mm}, height={61.5mm}, scaled x ticks={false}, xlabel={$\theta$ [º]}, x tick style={color={rgb,1:red,0.0;green,0.0;blue,0.0}, opacity={1.0}}, x tick label style={color={rgb,1:red,0.0;green,0.0;blue,0.0}, opacity={1.0}, rotate={0}}, xlabel style={at={(ticklabel cs:0.5)}, anchor=near ticklabel, at={{(ticklabel cs:0.5)}}, anchor={near ticklabel}, font={{\fontsize{9 pt}{11.700000000000001 pt}\selectfont}}, color={rgb,1:red,0.0;green,0.0;blue,0.0}, draw opacity={1.0}, rotate={0.0}}, xmajorgrids={true}, xmin={-2.700000000000003}, xmax={92.7}, xticklabels={{$0$,$20$,$40$,$60$,$80$}}, xtick={{0.0,20.0,40.0,60.0,80.0}}, xtick align={inside}, xticklabel style={font={{\fontsize{8 pt}{10.4 pt}\selectfont}}, color={rgb,1:red,0.0;green,0.0;blue,0.0}, draw opacity={1.0}, rotate={0.0}}, x grid style={color={rgb,1:red,0.0;green,0.0;blue,0.0}, draw opacity={0.1}, line width={0.5}, solid}, axis x line*={left}, x axis line style={color={rgb,1:red,0.0;green,0.0;blue,0.0}, draw opacity={1.0}, line width={1}, solid}, scaled y ticks={false}, ylabel={SER}, y tick style={color={rgb,1:red,0.0;green,0.0;blue,0.0}, opacity={1.0}}, y tick label style={color={rgb,1:red,0.0;green,0.0;blue,0.0}, opacity={1.0}, rotate={0}}, ylabel style={at={(ticklabel cs:0.5)}, anchor=near ticklabel, at={{(ticklabel cs:0.5)}}, anchor={near ticklabel}, font={{\fontsize{9 pt}{11.700000000000001 pt}\selectfont}}, color={rgb,1:red,0.0;green,0.0;blue,0.0}, draw opacity={1.0}, rotate={0.0}}, ymode={log}, log basis y={10}, ymajorgrids={true}, ymin={0.000275}, ymax={6.0}, yticklabels={{$10^{-3}$,$10^{-2}$,$10^{-1}$,$10^{0}$}}, ytick={{0.001,0.01,0.1,1.0}}, ytick align={inside}, yticklabel style={font={{\fontsize{8 pt}{10.4 pt}\selectfont}}, color={rgb,1:red,0.0;green,0.0;blue,0.0}, draw opacity={1.0}, rotate={0.0}}, y grid style={color={rgb,1:red,0.0;green,0.0;blue,0.0}, draw opacity={0.1}, line width={0.5}, solid}, axis y line*={left}, y axis line style={color={rgb,1:red,0.0;green,0.0;blue,0.0}, draw opacity={1.0}, line width={1}, solid}, colorbar={false}]
    \addplot[color={rgb,1:red,0.0;green,0.6056;blue,0.9787}, name path={475}, draw opacity={1.0}, line width={1.5}, solid, mark={*}, mark size={1.875 pt}, mark repeat={1}, mark options={color={rgb,1:red,0.0;green,0.0;blue,0.0}, draw opacity={1.0}, fill={rgb,1:red,0.0;green,0.6056;blue,0.9787}, fill opacity={1.0}, line width={0.75}, rotate={0}, solid}]
        table[row sep={\\}]
        {
            \\
            0.0  0.101590015900159  \\
            10.0  0.08853088530885307  \\
            20.0  0.09275292752927532  \\
            29.999999999999996  0.09516795167951679  \\
            40.0  0.1104061040610406  \\
            50.0  0.10922409224092242  \\
            59.99999999999999  0.12117421174211743  \\
            70.0  0.15385153851538522  \\
            80.0  0.20283202832028321  \\
            90.0  0.2640336403364033  \\
        }
        ;
    \addlegendentry {NC-M}
    \addplot[color={rgb,1:red,0.0;green,0.6056;blue,0.9787}, name path={476}, draw opacity={1.0}, line width={1.5}, solid, mark={square*}, mark size={1.875 pt}, mark repeat={1}, mark options={color={rgb,1:red,0.0;green,0.0;blue,0.0}, draw opacity={1.0}, fill={rgb,1:red,0.0;green,0.6056;blue,0.9787}, fill opacity={1.0}, line width={0.75}, rotate={0}, solid}]
        table[row sep={\\}]
        {
            \\
            0.0  0.0010120101201011923  \\
            10.0  0.0004640046400463856  \\
            20.0  0.0004000040000399922  \\
            29.999999999999996  0.00044700447004469477  \\
            40.0  0.0010110101101010922  \\
            50.0  0.0007350073500734873  \\
            59.99999999999999  0.0006600066000659877  \\
            70.0  0.0015810158101581078  \\
            80.0  0.004227042270422706  \\
            90.0  0.012329123291232908  \\
        }
        ;
    \addlegendentry {C-M}
    \addplot[color={rgb,1:red,0.8889;green,0.4356;blue,0.2781}, name path={477}, draw opacity={1.0}, line width={1.5}, dashed, mark={*}, mark size={1.875 pt}, mark repeat={1}, mark options={color={rgb,1:red,0.0;green,0.0;blue,0.0}, draw opacity={1.0}, fill={rgb,1:red,0.8889;green,0.4356;blue,0.2781}, fill opacity={1.0}, line width={0.75}, rotate={0}, solid}]
        table[row sep={\\}]
        {
            \\
            0.0  0.0928119281192812  \\
            10.0  0.09553895538955387  \\
            20.0  0.0962809628096281  \\
            29.999999999999996  0.098489984899849  \\
            40.0  0.10867608676086761  \\
            50.0  0.1116121161211612  \\
            59.99999999999999  0.13854638546385462  \\
            70.0  0.17351373513735135  \\
            80.0  0.31883718837188363  \\
            90.0  0.4364793647936479  \\
        }
        ;
    \addlegendentry {NC-MM}
    \addplot[color={rgb,1:red,0.8889;green,0.4356;blue,0.2781}, name path={478}, draw opacity={1.0}, line width={1.5}, dashed, mark={square*}, mark size={1.875 pt}, mark repeat={1}, mark options={color={rgb,1:red,0.0;green,0.0;blue,0.0}, draw opacity={1.0}, fill={rgb,1:red,0.8889;green,0.4356;blue,0.2781}, fill opacity={1.0}, line width={0.75}, rotate={0}, solid}]
        table[row sep={\\}]
        {
            \\
            0.0  0.003790037900379006  \\
            10.0  0.00844008440084402  \\
            20.0  0.05693756937569375  \\
            29.999999999999996  0.23465434654346543  \\
            40.0  0.44613246132461326  \\
            50.0  0.5667746677466775  \\
            59.99999999999999  0.6510085100851007  \\
            70.0  0.6995079950799508  \\
            80.0  0.60149901499015  \\
            90.0  0.3833768337683376  \\
        }
        ;
    \addlegendentry {C-MM}
    \addplot[color={rgb,1:red,0.2422;green,0.6433;blue,0.3044}, name path={479}, draw opacity={1.0}, line width={1.5}, dotted, mark={*}, mark size={1.875 pt}, mark repeat={1}, mark options={color={rgb,1:red,0.0;green,0.0;blue,0.0}, draw opacity={1.0}, fill={rgb,1:red,0.2422;green,0.6433;blue,0.3044}, fill opacity={1.0}, line width={0.75}, rotate={0}, solid}]
        table[row sep={\\}]
        {
            \\
            0.0  0.0879288792887929  \\
            10.0  0.1075990759907599  \\
            20.0  0.10383803838038379  \\
            29.999999999999996  0.10376803768037682  \\
            40.0  0.10517105171051708  \\
            50.0  0.1143541435414354  \\
            59.99999999999999  0.12847228472284725  \\
            70.0  0.15318153181531816  \\
            80.0  0.20067000670006702  \\
            90.0  0.20056800568005687  \\
        }
        ;
    \addlegendentry {NC-U}
    \addplot[color={rgb,1:red,0.2422;green,0.6433;blue,0.3044}, name path={480}, draw opacity={1.0}, line width={1.5}, dotted, mark={square*}, mark size={1.875 pt}, mark repeat={1}, mark options={color={rgb,1:red,0.0;green,0.0;blue,0.0}, draw opacity={1.0}, fill={rgb,1:red,0.2422;green,0.6433;blue,0.3044}, fill opacity={1.0}, line width={0.75}, rotate={0}, solid}]
        table[row sep={\\}]
        {
            \\
            0.0  0.00047400474004738826  \\
            10.0  0.0010750107501074932  \\
            20.0  0.0010940109401093912  \\
            29.999999999999996  0.0007700077000769948  \\
            40.0  0.0011370113701136904  \\
            50.0  0.0009360093600935815  \\
            59.99999999999999  0.0012240122401223964  \\
            70.0  0.001353013530135294  \\
            80.0  0.003296032960329594  \\
            90.0  0.003321033210332099  \\
        }
        ;
    \addlegendentry {C-U}
\end{axis}
\end{tikzpicture}
    				}
    				\caption{Error probability of the matched (M) and mismatched (MM), coherent (C) and noncoherent (NC), and uncoupled (U) detectors at azimuth angles from $\qty{0}{\degree}$ to $\qty{90}{\degree}$. The array is $D=\qty{0.5}{\m}$ and $N=128$.}
    				\label{fig:ser_coupled_theta}
                \end{figure}

                Although the previous results are valid in general, the exact performance is not independent of the user location~\cite{damico_holographic_2024}.
                In~\cref{fig:ser_coupled_theta} it can be seen that, at end-fire, the coherent detector \acrshort{ser} is lower than the noncoherent one, even for $N=128$.
                Of course, if the number of antennas keeps increasing, the performance of the former will reduce whereas the latter's will improve, as discussed previously.
                Furthermore, the figure suggests that coupling primarily affects the response phase while leaving the power almost unchanged.
                Since the noncoherent detector relies solely on power, its performance remains unaffected by the mismatch.

            \section{Conclusions}
                In this paper, we examined the problem of inaccurate modeling of mutual coupling in a receiver with closely spaced antennas.
                Specifically, we analyzed the uplink of a communication system where the \acrshort{bs} neglects the effects of mutual coupling, and compared the resulting error probability with that obtained using the correct model.
                Our analysis employed both coherent and noncoherent decoding schemes, revealing that noncoherent decoding is more robust to coupling mismatch.
                This is because the noncoherent receiver does not exploit phase information but only power, remaining less sensitive to mutual coupling mismatches.
                
                The aforementioned results suggest that, in situations where mutual coupling cannot be accurately modeled, it may be beneficial to ignore phase information and instead rely on instantaneous channel energy detection~\cite{jing_design_2016}.
                In general, this approach provides a performance level between that of the coherent detector and the average channel energy detector considered in this paper.
                Furthermore, the effects of mutual coupling on both the power and phase of the array response should be further explored, as well as the implications of wavefront curvature on the coupling behavior.
        
        \bibliography{references}
\end{document}